# Helical Nanomachines for Fast Mechanical Mapping of Heterogeneous Environments

Arijit Ghosh[†] and Ambarish Ghosh[†,‡,¶,*]

† Department of Electrical Communication Engineering, Indian Institute of Science, Bangalore
‡ Department of Physics, Indian Institute of Science, Bangalore
¶ Centre for Nano Science and Engineering, Indian Institute of Science, Bangalore

*Corresponding author e-mail: ambarish@ece.iisc.ernet.in

**ABSTRACT**:

Artificial micro and nano machines have been envisioned and demonstrated as potential candidates for variety of applications, ranging from targeted drug or gene delivery, cell manipulation, environmental sensing and many more. Here, we demonstrate the application of helical nanomachines that can measure and map the local rheological properties of a complex heterogeneous environment. The position of the helical nanomachine was controlled precisely using magnetic fields, while the instantaneous orientation provided an estimation of the viscosity of the surrounding medium with high spatial and temporal accuracy. Apart from providing viscosity estimates in purely viscous and viscoelastic media with shear rate independent viscosity (Boger fluids), their motion was also found to be extremely sensitive to fluid elasticity. Taken together we report a promising new technique of mapping the rheological properties of a complex fluidic environment by helical nanomachines with high spatial and temporal resolutions, a functionality that goes beyond the capabilities of existing passive and active microrheological methods.

**KEYWORDS**: Nanomachines, Active Microrheology, Viscosity, Viscoelasticity, Helical Propulsion, Magnetic Nanoswimmers

Artificial micro/nanomachines have been a topic of active research over the last few years primarily due to their potential as noninvasive tools for therapeutics and diagnostics [1-2]. They have already been rendered suitable for in vivo [3-4] applications, shown as carriers for microscale objects [5-6], drugs [7-8], genes [9], and have been used for various lab on chip sensing applications [10-11]. There are different ways of powering and controlling nanomachines using magnetic fields [12], ultrasound [13-14], light [15-16] or chemical redox reactions [17], of which helical magnetic nanomachines are particularly relevant to the applications demonstrated here. The machinery relies on corkscrew type motion of small ferromagnetic helices induced by a rotating magnetic field [18-19], which have been maneuvered in various biological environments, such as human blood [20], serum/saliva [6, 21], mucin [22], peritoneal cavity [3] etc. The helical shape of these machines is critically important in converting the magnetically induced rotation to linear translation, similar to the way various helically flagellated microorganisms are self-propelled [23] at low Reynolds numbers.

Although these machines have been successfully maneuvered in different environments, the possibility of using these machines as measurement probes for the local physical and/or chemical environment has been relatively [24] unexplored. In this work, we report for the very first time, the application of helical magnetic nanomachines as probes for measuring and mapping the local rheological properties of an environment with complex heterogeneities. The experiments and analysis described here not only empowers the helical nanomachines with a new functionality, but also shows measurements with significantly higher spatial and temporal resolution compared to conventional microrheological techniques. The standard method of microrheology involves the measurement of the position fluctuations of passive probe particles dispersed in a sample of interest, using video microscopy [25-26], light scattering [27] etc, from where it is possible to extract the rheological parameters of the surrounding medium. The measurement times for passive microrheology are typically long (~100 seconds or more), in comparison to active [28-30] methods, where the rheological information is extracted from the dynamical response of the probe particle under the action of externally applied optical [31] or magnetic [32-33] forces or torques. Both these methods are suitable for making highly localized mechanical measurements, but the control on the spatial location of the probe is either absent, or requires invasive techniques of micromanipulation, such as an optical tweezer.

In the method described here, we maneuvered the probe particle (here, the helical nanomachine) with micron scale resolution in fluidic environments [18], while simultaneously imaging its position and orientation [31] to obtain the local rheological information in real time. The method of manipulation, based on small, homogeneous, rotating magnetic fields is non-invasive and compatible with almost any type of sample, including living systems. The paper is organized as follows. In the first part of the paper, we consider an elongated chiral (here, helical) nanostructure under external torque. We show how the different dynamical configurations [34] of the nanostructure depend on the relative strength of the frictional and the driving torques, and simply by imaging the dynamics how one can estimate the viscosity of the local medium (assumed to be frequency independent) surrounding the structure. As a proof-of concept we measured the spatiotemporal variation of viscosity in a microfluidic chamber containing two fluids that mix with each other during the course of the experiment. We found this technique to be significantly faster and more accurate than conventional microrheological methods, and only limited by the inherent thermal fluctuations of the surrounding medium. As an application of this technique, we investigated the motion, in particular speed of helical machines in a viscoelastic medium. This was obtained by driving the nanostructure across a region with varying degree of concentration of an elastic material. We used the methodology described in the first part to estimate the viscosity and thereby concentrations of the elastic material locally. This was followed by relating the speed of the machine to concentration of elastic material, allowing a self-referenced measurement of helical propulsion as a function of elastic relaxation timescale. Taken together, we report on a promising new development in mapping complex heterogeneous mechanical environments with fast and accurate quantitative measurements.

### Fabrication and actuation of the helical nanoprobes:

The experimental system of nanoscale helices (Fig. 1a), also referred to as the magnetic nanopropellers in the literature, was fabricated using a physical vapor deposition technique called GLancing Angle Deposition (GLAD) [35], and actuated with rotating magnetic fields generated in a triaxial Helmholtz coil. Some of the details about the experimental system have been published before [18, 34], with some additional information given in the methods section. We have described the actuation and imaging system used for these measurements in section I of the supporting information.

### Dynamics of helical nanoprobes – An estimate of local viscosity:

When a ferromagnetic elongated structure like a nanorod or a nanohelix is subjected to a rotating magnetic field, the dynamics of the structure is governed by the tendency of the magnetic moment to be pointed along the direction of the externally applied field. In the calculations and experiments presented here, we assume the moment of the nanostructures to be permanent and have neglected any possible effects arising out of induced moments (superparamagnetism) or demagnetization (coercivity assumed to be higher than the applied fields). As shown in Fig. 1b, the magnetization was assumed to be of strength

*M* at an arbitrary angle $\theta_m$ to the short axis of the helix. The strength and frequency of the rotating magnetic field was $B_0$ and $\Omega_B$ respectively. This problem was considered theoretically in one of our earlier papers [36] (see section II of the supporting information), where the dynamical equations were solved for an elongated nanorod. The relevant results of our calculations are presented here: (i) the motion was considered to be inertia less (Reynolds number << 1) in an unbound Newtonian fluid and Euler equations were used to solve the dynamics. (ii) At very low values of $\Omega_B$, the rod tumbles, in which it rotates about its geometric short axis. Tumbling continues till a critical frequency $\Omega_1$, given by: $\Omega_1 = \frac{MB}{\gamma_s}\sin\theta_m$ where $\gamma_s$ is the rotational drag coefficient about the short axis. (iii) Above the critical frequency $\Omega_1$, the rod starts precessing about the direction perpendicular to the plane of the rotating field, and the angle of precession, $\alpha_p$ (see Fig. 1B), was derived to be $\alpha_p = \sin^{-1}(\Omega_1/\Omega_B)$. (iv) Only frequencies at which the moment rotates in sync with the field, i.e. below the step-out frequency, is relevant to the work reported here.

The scheme of measurement presented here relies on the dependence of precession angle $\alpha_p$ on the drag coefficient $\gamma_s$, which in turn is directly proportional to the viscosity ($\eta$) of the local surrounding and geometry dependent factor $f_g$.

$$\alpha_p = \sin^{-1}\left[\frac{MB_0\sin(\theta_m)}{\gamma_s\Omega_B}\right] = \sin^{-1}\left[\frac{MB_0\sin(\theta_m)}{\eta f_g\Omega_B}\right] \quad (1)$$

Therefore, if the dimensions and magnetic moment of the helix are known, it is possible to estimate the viscosity ($\eta$) of its surrounding from its angle of precession $\alpha_p$ (see section III of supporting information for a detailed method of measuring precession angle). If the viscosity $\eta$ of the surrounding medium is changed either by inducing a rheological transition in the medium or by transporting the object to a location of different mechanical property, the precession angle would change, which would thereby provide a quantitative estimate of the "local" surrounding. The measurement time would be determined by the time it takes to measure the precession angle of the helix, which in turn depends on the frequency of the magnetic field, and therefore can be quite fast. This is the central idea of the scheme used to measure the local viscosities. In order to increase the dynamic range of this technique, one can increase or decrease the magnetic field $B_0$, which will enable the measurement of a widely different value of $\eta$ using the same value of $\alpha_p$. In the supporting information (section IV) we have discussed the variation of the precession angle as a function of viscosity and magnetic field, and ways to improve the dynamic range of this measurement scheme.

Equation 1 was originally derived for elongated objects in which the rotational and the translational degrees of freedom are not coupled. Therefore, the formula is only approximately (not exactly) correct for a helical geometry, as the mobility tensor contains extra terms originating from the coupling. As a result, the axis of a finite sized helix can deviate from the principal axes of rotation, and the rotational drag coefficients have to be modified with contributions from the coupling terms of the mobility tensor. This issue was first considered with approximate calculations by Morozov et. al. [37], who estimated the misalignment of the axes to be negligible for typical experimental geometries. A more detailed theoretical analysis was recently published by Fu et. al. [38]. To summarize these theoretical analyses, it is indeed justified to split the dynamics of a helix into two problems: (i) precessional motion of a rod governed by equation 1 with an effective $\gamma_s$ that slightly differs from that of a rod, and (ii) translation motion of a helix due to rotation about its long axis, that depends on the angle of precession and an effective hydrodynamic pitch. To check it experimentally, we have collected data with various magnetic field parameters. The results are shown in Fig. 1c, where $\alpha_p$ is plotted as a function of $\Omega_1/\Omega_B$ in water at magnetic field magnitudes 10 G and 20 G and in 0.3% methyl cellulose (MC) gel at 30 G and 60 G. We have included two representative movies [SM1 and SM2] in the Supporting Information, in which the same helix was rotated in 0.3% methyl cellulose at two different rotation frequencies 10 Hz ($\alpha_p$ = 56°) and 20 Hz ($\alpha_p$ = 30°) at the same field magnitude of 30 G. The measurements indicate clearly that the precession angle of the helical machines both in viscous and viscoelastic media at different field magnitudes can be well described with the form shown in equation 1, $\alpha_p = \sin^{-1}(\Omega_1/\Omega_B)$. It is important to note that the viscosity of the MC gel at such low concentrations remains unaltered at low shear rates (as in these experiments) and therefore it is not surprising that the equation derived for a purely Newtonian fluid was applicable here.

**Viscosity mapping in heterogeneous media:**

The experimental procedure, conducted in a microfluidic chamber, is shown schematically in Fig. 2a, where a concentration gradient of a test fluid was created by dispersing the test fluid and a reference fluid from the two inlets of the device. The reference fluid, typically water or some other medium with known viscosity ($\eta_{ref}$) containing large number of helical nanoprobes provided a reference for the measurement. Within the time frame of our experiments which varied between few minutes to an hour, the concentration gradient between the fluids could still be visible, although there was some mixing around the centre of the microfluidic chamber. A photograph of a device containing two fluids is shown in Fig. 2b. The propellers were maneuvered from the reference medium to the other fluid of unknown viscosity ($\eta_{meas}$), and analyzing their images provided a measurement of the precession angles ($\alpha_{p,ref}$ and $\alpha_{p,meas}$) in the two media and across the gradient. The viscosities could be estimated from the values of the magnetic field strengths $B_{ref}$ and $B_{meas}$, and frequency $\Omega_{ref}$ and $\Omega_{meas}$ used for the reference and the test media respectively. The viscosity of the test medium was obtained using $\frac{\eta_{meas}}{\eta_{ref}} = \frac{B_{meas}}{B_{ref}} \frac{\Omega_{ref}}{\Omega_{meas}} \frac{\sin\alpha_{p,ref}}{\sin\alpha_{p,meas}}$

Although the measurement of $\alpha_p$ was most sensitive when it was slightly less than π/2, the speed of the propellers in the tumbling state was too low, i.e. $v_p \sim 0$. Accordingly, we carried out the experiments keeping $\alpha_p \sim 30° - 50°$, while adjusting $B_{meas}$ and $\Omega_{meas}$ as the helices were moved from one end of the chamber to another within a reasonable time. A key point to note is that the experimental procedure is self-referenced and therefore does not require prior measurement of the magnetic moments or the friction coefficients.

In Fig. 2c, we show results of the viscosity measurements as the nanohelices were taken from de-ionized water across the concentration gradient to a different fluid. We present data for two experiments with the other fluid being 55% Glycerol solution (top graphs) and 30% w/w aqueous dextrose solution (bottom graphs) respectively. Typically, we travelled few mm from the starting point, and measured the local viscosity as a function of position. The measurement parameters are given in the methods section. Note the final positions of the helices for the cases shown in Fig. 2c were somewhere deep in between the two pure phases of the liquids, and therefore it is not surprising that the estimated viscosity of 5 cP and 2.33 cP were smaller than the bulk viscosity of the test samples, 8.7 cP and 10 cP respectively.

Due to mixing of the miscible fluids inside the microfluidic chamber, local viscosities at different points would slowly change as a function of time. To verify whether the local viscosities measured by the nanohelices were indeed accurate, an alternative scheme was used in which we observed the position fluctuations [39] of the nanohelices in the absence of the magnetic field at the starting and end locations. The passive fluctuations of the helices were inversely related to the local viscosity of a location, and therefore measuring the diffusivities of the helices along their long and short axes provided alternate estimates of the viscosity. These measurements were also self-referenced since we measured the ratio of the diffusivities at starting and end locations using the same helix, giving ratio of the viscosities at the two end points. The results of the passive measurements of the ratio of the diffusion constants matched very well with the estimates obtained from the precession angle measurements at the end point, as marked clearly in Fig. 2c by star symbols, which validates the quantitative accuracy of this technique.

The measurements were carried out with an averaging time of 10 to 15 seconds while the helices were moving at a speed of around 1 µm/s, implying the spatial resolution of the measurement was around 10 to 15 µm. The same fluidic systems were also used to measure the temporal evolution of the viscosity as the fluids mixed with time. At the start of the measurement, the helices were brought close to the test fluid and were kept precessing. By periodically reversing the sense of rotation of the magnetic field, it was possible to alter the direction of motion and thereby confine the probes in a region within tens of microns. The estimate of the viscosity as a function of time is shown in Fig. 2B, along with the estimate of the viscosity obtained from passive measurements at the end of the measurements. The measurement times shown here were around 15 seconds. We believe the averaging times can be easily reduced by two orders of magnitude, using higher rotation frequencies and/or in fluids of higher viscosities (see Fig. 3b). The range of viscosities measured during the course of our experiments, including the measurements presented in the next section, was between 1 and 15 cP. This was related to the dimensions (hence, drag) of the nanomachines as well as maximum magnetic torque available to drive the machines. With the present technique of nanofabrication, we are able to drive machines of similar dimensions in media with viscosities as high as almost a hundred cP.

There are several sources of error in estimating the viscosity, which largely originates in estimating the orientation of the helix from the sequence of images. This can depend on the details of the imaging system, e.g. the pixel size and speed of the camera, and also how stable the applied magnetic fields are, which in turn would depend on the current amplifiers used to drive the coils. Errors like this may be reduced by engineering better setups; however, these sources only form a small fraction of the uncertainty that was experimentally measured. As we discuss here, the accuracy of the viscosity estimation with this method is primarily limited by the orientation fluctuations of the probe due to inherent thermal noise of the surrounding environment. The fluctuating orientation angle can be measured with an accuracy that depends on temperature, as well as measurement time and speed (frames per second). To estimate the error arising from thermal noise, we simulated the motion of the probe based on our earlier work [39] on thermal fluctuations of the nanohelices. We obtained a time series of the orientation angle, and this was done for a range of viscosities. Subsequently the data was sampled at 24 measurements of angles per second for a total time of 2 seconds. The error in measuring the precession angle was obtained by fitting the orientation angle to a sinusoid, and the simulation results are shown in Fig. 3a. For comparison, we considered the experimental data of MC gel (data shown in Fig. 4b) at different concentrations. These experiments were performed under similar conditions (24 frames per second and duration 2 seconds) as the simulations. As expected, the error reduced as the viscosity of the medium was increased. The simulated and experimentally measured errors were reasonably close to each other in the range of viscosities investigated here, implying that the accuracy in our method is primarily limited by the inherent thermal fluctuations of the system with relatively small contributions from all other sources of error. For example, considering water as the reference fluid, the average experimental uncertainty was almost 14% for the measurement of 1 cP viscosity, while the uncertainty from the simulated data was about 11%. At viscosities of around 12 cP, the experimental uncertainty reduced to about 3%, while the simulated error was about 4%. More details on error estimates with and without self-referencing techniques can be found in section V of the supporting information. Our calculations show that the relative error in measuring viscosity can be related to the error in measuring the precession angle as $\frac{d\eta}{\eta} = -\cot\alpha_p \, d\alpha_p$. We have plotted this quantity in Fig 3b for the same measurements shown in Fig 3a as a function of the time of measurement. We find the error to be around 5% for measuring a viscosity of around 14 cP. In order to highlight the speed of our technique, we have also shown in the same graph, the errors in obtaining the viscosity estimate by passive microrheology technique using a 1 µm diameter miocrosphere. It can be seen that for a measurement time as low as 0.2 seconds, our approach can be used to measure the viscosity with sufficient accuracy, while it is almost impossible to get any realistic estimate using passive micorheology. It may be noted that in this case the helix was rotated at a speed of 5 Hz and thus the smallest time shown here corresponds to one rotation cycle of the helix. With higher rotation speeds, the time of measurement can be reduced further which shows the ability of this technique to extract information in almost real time. Thus to our knowledge this is the fastest viscosity measurement technique available till date.

In situations where a reference fluid is not available, it becomes necessary (according to equation 1) to know the drag coefficient and magnetization angle ($\theta_M$) of the helical nanostructure, which could introduce additional sources of error and would depend on the details of the measurement and fabrication scheme. One way to avoid these errors will be to make a passive measurement at a certain location to estimate the local viscosity and make a reference measurement, and subsequently follow the method described in detail before.

Since our viscosity measurements were carried out in purely Newtonian and Boger fluids (viscosity independent of shear rate), it is natural to wonder whether one could use the same technique for shear thinning/thickening fluids as well. This may be achieved by driving the nanomachine at different magnetic field frequencies, and hence different shear rates, and subsequently relating the precession angle to a shear rate dependent viscosity by modifying equation 1 appropriately. While this may provide mechanical information at many time scales, the measurement process will sequential and therefore slow.

### Investigating the effect of elasticity on the speed of helical propulsion

Motion of helical swimmers in complex viscoelastic media has been a topic of active research [40-42] in the recent past. Experiments revealed both decrease [41] as well as increase [40] of the speed of helical propulsion in elastic media, which as shown recently [41] is a complicated function of the helical geometry, as well as the ratio of the actuation to the relaxation time scales of the elastic medium. In the experiments presented here, we chose methyl cellulose gel as the model system and investigated the speed of the nanohelices as a function of concentration of the elastic medium. Although methyl cellulose is known to be shear thinning, it can be considered to follow a Maxwell model for linear viscoelasticity at shear rates typical (5 to 40 rads$^{-1}$) of our experiments (see Section VI of supporting information for a detailed rheological characterization of the gel). Note the representative length scales (~ 10 nm) [43] for these samples such as the mean spacing between the polymer meshes were significantly less than the sizes of the helices (~ 1 µm), and therefore the elastic response of the system was indeed probed as a whole.

As shown in the schematic of Fig. 4a, the procedure of the experiments was similar to what has been described before, where the nanohelix was maneuvered from water to the other side of the microfluidic device containing a solution of 0.5% methyl cellulose. As the helix was moved across the boundary it was alternately driven in the helical propulsion mode ($\Omega_B$ = 25 Hz) and the precessional mode ($\Omega_B$ = 5 Hz). The strength of the magnetic field was varied such that the precession angles were around 30º - 50º, and the driving torque during the propulsion was always greater than the frictional torque. The idea was to measure the pitch and the viscosity (from precession angle) at approximately the same location, where the hydrodynamic pitch is the distance moved by the helix for one complete rotation about its long axis. In Fig. 4b, a plot of the hydrodynamic pitch has been shown as a function of the distance across the boundary along with the viscosity measurement obtained from precession angle measurements. Each point here denotes pitch over a measurement distance of 100 µm. Very interestingly, within a small distance (< 900 µm) from the starting point (within DI water) a large (~ 30%) reduction in speed (hydrodynamic pitch) was observed for a negligible change in the viscosity. The speed reduction of the helices can be clearly seen in the movies (SM3 and SM4) available in the supporting information. As the helical probe was moved further into the gel, the pitch increased to values almost similar to that in water. The measured values of viscosity, on the other hand, increased monotonically as the propellers moved from water towards methyl cellulose, similar to what we obtained in our spatial viscosity mapping experiments discussed in Fig. 2.

In Fig. 4c, we show the variation of the pitch of the helices as a function of gel concentration, where the concentration was obtained from local viscosity estimates, which in turn was obtained from precession angle measurements. We show data for different helices, including the one shown in Fig. 4b. These data were collected in different chambers, of which the data shown in circles had 0.3% gel at the beginning of the experiment, while the other two datasets (shown as squares and triangles) contained 0.5% gel. Irrespective of the concentration of the gel used in one end of the chamber, the pitch could be measured at a concentration as low as 0.01%, which corresponds to a viscosity of 1.4 cp, below which the method of relying concentration from local viscosity measurement is not accurate. All the different datasets clearly show a common general trend, where the hydrodynamic pitch increases by around 30% as the gel concentration increased from 0.01% (relaxation time 0.31 seconds) to 0.45% (relaxation time 0.5 seconds). The exact absolute values of the pitch varied between the helices, which arose due to their structural differences, and would also happen in a Newtonian fluid. However, the strong and similar sensitivity of the pitch on the relaxation time of the elastic medium is in agreement with recent analysis of helical propulsion of biological swimmers in viscoelastic media [41], and qualitatively similar to a recent measurement [42] of artificial helical flagella in gels. The key point that must be stressed here is that our measurements were inherently self-referenced, in which the same propeller could be used to do the measurements of viscosity and speed at the different concentrations of the gel, and as far as we know there is no other technique that can achieve the same. In standard experiments, different helical swimmers are studied in different chambers containing varying concentrations of viscoelastic fluids, and therefore the measurements are subject to variability within the swimmers, a disadvantage that is completely circumvented by our method.

Magnetic nanohelices, which can be rendered biocompatible [20], can be either driven precisely [18] in a certain path or moved autonomously [45], have here been used as active probes for microrhelogical mapping of complex heterogeneous environments. These helical machines provide several advantages over existing techniques, such as high spatial accuracy and speed of measurement. As the technique only requires imaging the helices, it may be combined with other applications previously demonstrated with helical magnetic nanomachines, such as cargo delivery [6]. Although the present experiments were performed with a single nanohelix under observation, it may be possible to extend these experiments with two or more helices independently positioned [46] at different locations of the microfluidic chamber, or in applications requiring autonomous [45] sampling along multiple directions in a certain region. We envision this technique to be useful in scenarios that require fast mechanical changes such as gelation studies [47], or materials with complex mechanical heterogeneities, which could even be a biological cell [48]. Also the very large sensitivity (30% reduction for the presence of 50 ppm of MC gel) of the pitch to the lower concentrations of the gel shows the possible use of the nanohelices for the detection and quantification of minute quantities of elastic impurities. For such measurements, it will be necessary to make a prior measurement of the speed as a function of concentration of the elastic medium. A simultaneous measurement of viscosity is not necessary since a look-up table for speed vs concentration has been made already. As we demonstrate here, this technique would allow the investigation of helical swimming [40] in viscoelastic media in a completely self-referenced manner, which would be of great fundamental interest. We are sure with further work [49], this method could be extended to other non-Newtonian viscoelastic fluids, and also estimate the frequency dependent elastic parameters of more complex rheological environments. Finally, it will be interesting to explore biological environments with this rheological tool, which would probably be less susceptible to protein fouling effects compared to nanomotors based on catalytic reactions [50]. However, the dimensions of the magnetic nanohelices in relation to the characteristic length scale of the ingredients (polymeric mesh [44] or cellular suspensions [20]) would play an important role in understanding their dynamics in biofluids. In view of the current interest and advances in the development of intelligent nanomotors [51] for biological [2, 48] applications, this technique may be extended to other [13] nanomotor designs as well.

## METHODS

*Fabrication of magnetic nanohelices:*

The magnetic nanohelices were fabricated using a physical vapor deposition technique called GLancing Angle Deposition. At first a monolayer layer of polystyrene beads (700 nm diameter) were formed using Langmuir Blodgett deposition technique. The monolayer was subsequently etched using plasma etching (Harrick Plasma) for 12 minutes in medium RF power to obtain beads of reduced sizes of around 550 nm [52]. This layer acted as the seed layer for the growth of the helices and the width of the helix was controlled by controlling the diameter of the beads. Silicon dioxide was then evaporated at an extreme angle by electron beam evaporation, while the substrate was manipulated in a controlled manner. As a result, helical structures of controlled shape and size could be fabricated. The nanostructures were taken out using sonication in water and laid down on Si wafers. They were rendered magnetic by depositing a thin layer of cobalt (≈ 50 nm) on one side of the helix, with a 10 nm layer of chromium to improve the adhesion.

*Fabrication of the microfluidic device:*

Microfluidic devices were fabricated using thin (100 µm) double sided tapes, glass slides and cover slips. As shown in the schematic of Fig. 3A, double sided sticky tapes were used to form channels in between a glass slide and a microscope cover glass. Generally, two inlets and two outlets were kept on the opposite sides. Thus when two different fluids were injected through the different inlets (using surface tension forces) a two fluid boundary gets formed between the two glass walls. After the formation of the boundary, all the gaps were sealed using glue. In order to visually identify the boundary, which was stable for over an hour, sometimes one of the fluids was colored using a small quantity of methylene blue dye and filtered in 0.2 µm syringe filter. The helices were dispersed in the reference fluid which was in general DI water.

*Measurement of spatial variation of viscosity:*

The naomachines were first dispersed in the fluid of known viscosity (generally DI water). One of the helices at the extreme end of the water side was then chosen for measurement. First a passive rheological measurement was performed, in which the translational diffusion coefficients of the helix [39] were measured in water. The helix was then moved across the two fluid boundaries and switched between the states of precession and propulsion. While precessing, the angle of precession was measured, which provided an estimate of the local viscosity of the surrounding (see equation 1 above). Once the measurement was done, the helix was propelled to a new place having different rheological property to carry out the viscosity measurement. The process continued until the helix reached somewhere deep inside the boundary. The magnetic field was then turned off and the diffusion coefficients measured. The ratio of the diffusion coefficients in water and at the end point provided an accurate alternate estimate of the local viscosity at the final point of measurement.

*Measurement of temporal variation of viscosity:*

The nanomachines were dispersed at the water end of the microfluidic device. To carry out the temporal measurements, a helix close to the boundary in the water end was chosen. Similar to the spatial measurements, the translational diffusion coefficients were measured, by imaging the helices in suspension, while no magnetic field was applied. The helix was kept in the precession configuration by moving back and forth (by alternatively applying a clockwise and an anticlockwise rotating field) and the value of $\alpha_p$ was measured continuously using high frame rate video microscopy. As the two fluids mixed, the local viscosity at the point of measurement increased. After around 15 to 20 minutes, the measurement was stopped and the diffusion coefficients of the helix were measured at this time. The ratio of the initial and final diffusion coefficients provided an accurate alternate measure of the viscosity at the final instant.

It may be noted that in both the spatial and temporal mapping experiments, the precession angle was kept at a measurable value (see section 2.3 above) by increasing the magnetic field value in steps, to probe the increasing local viscosity of the surrounding.

*Preparation of methyl cellulose solutions:*

Following standard recipes [53], at first one third of the required volume of DI water was heated to a temperature of around 90° C. Then the desired amount of methyl cellulose powder (%w/w) (Methocel 90 HG, purchased from Sigma Aldrich), was mixed thoroughly in the hot water using a magnetic stirrer till all the particles were wetted properly and the solution looked turbid. Then the heating was removed and icy cold water in the remaining proportion (two third of the required amount) was added to the solution, as the stirring was continued. The solution lost its turbidity and became transparent. The stirring was continued for one or two hours more and the solution was then refrigerated at 4° C overnight as the viscosity increased to the desired value. The viscoelastic properties of methyl cellulose (MC) solutions were characterized using passive microrheology (see section V of the supporting information).

*Simultaneous measurement of speed and viscosity:*

After their dispersion in the DI water end of the microfluidic cell, one of the helices was chosen at the water end. It was then moved across the boundary at a rotation speed of 25 Hz, while the propulsion speed was measured continuously, averaged over a length of around 100 µm. As the propeller was moving, it was brought to precessional motion using a magnetic field frequency of 5 Hz while keeping the magnetic field same. As the propellers reached a location with higher concentration of MC gel (therefore higher viscosity), to keep the precession angle measurable, we increased the strength of the magnetic field.

### Acknowledgements

The usage of the facilities in Micro and Nano Characterization Facility (MNCF, CeNSE) at IISc, and funding from the Department of Biotechnology, as well as Science and Engineering Research Board is gratefully acknowledged. This work is partially supported by the Ministry of Communication and Information Technology under a grant for the Centre of Excellence in Nanoelectronics, Phase II.

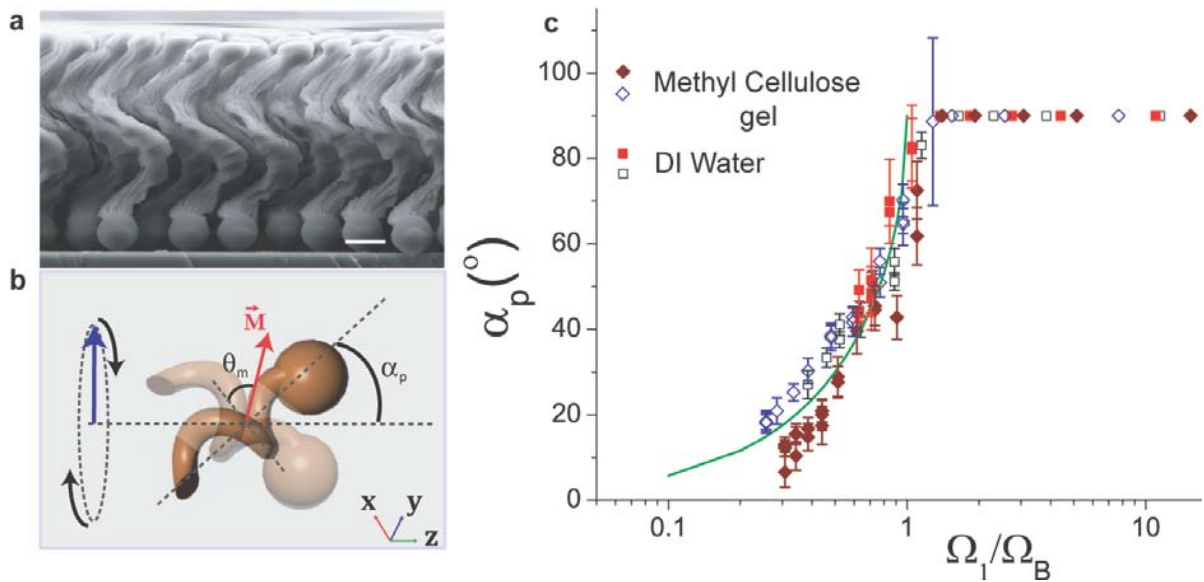

**Figure 1**: (a) SEM image of the system of magnetic nanohelices (scale bar equals 500 nm). (b) Schematic representation of a helical nanoscale probe with permanent magnetic moment at an angle $\theta_m$ to the short axis. Under the action of a magnetic field of magnitude $B_0$ and frequency $\Omega_B$ the helix precesses about the perpendicular (z direction) to the plane of the rotating magnetic field (x-y plane) with angle $\alpha_p$. (c) Plot of the angle of precession as a function of $\Omega_1/\Omega_B$ in purely viscous medium water (square symbols) and the viscoelastic gel 0.3% methylcellulose (diamond symbols). Representative movies are shown in the Supporting Information. While the open symbols correspond to a measurement at lower magnetic fields (10 G for water and 30 G for methyl cellulose), the solid symbols correspond to measurements at double the field magnitudes (20 G for water and 60 G for methyl cellulose). The green solid line corresponds to a plot of $\sin^{-1}(\Omega_1/\Omega_B)$.

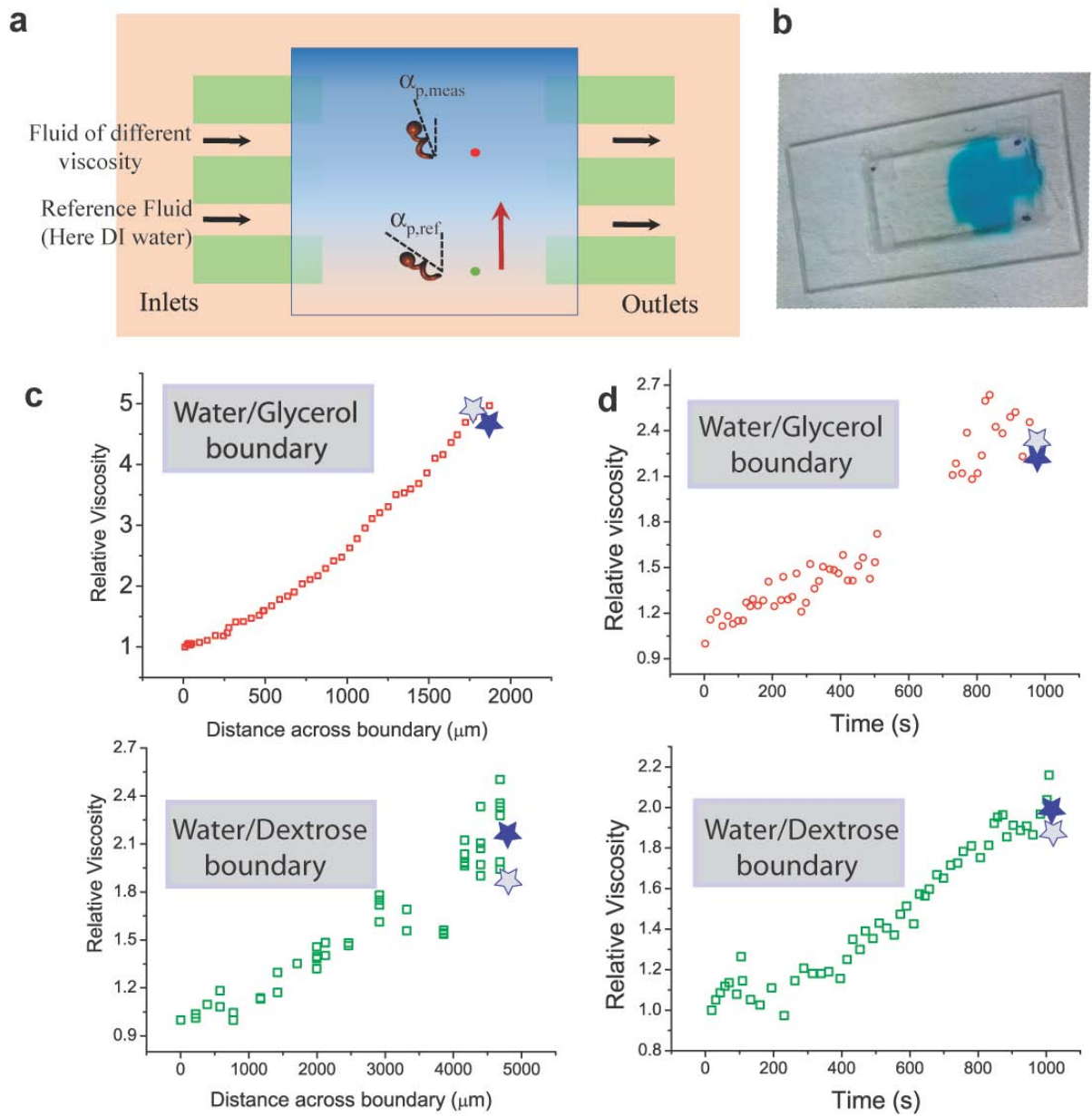

**Figure 2**: (a) Schematic of the microfluidic device. The chamber has two inlets and two outlets, with deionized (DI) water and a test fluid (blue color) flowing through. The propeller is driven from the water side towards the test fluid along the concentration gradient. Measurements of the precession angle at the reference point (green solid circle) and the measurement point (red solid circle), gives a measure of the relative viscosity at that particular point. (b) Photograph of the actual device where the two different fluids can be distinctly observed. One of the fluids was mixed with methylene blue dye for easier visualization. (c) Relative viscosity measured across the two fluid boundaries of glycerol-water (top) and dextrose-water (bottom) mixtures plotted as a function of the distance across the boundary. The estimates of the viscosity obtained from passive measurements have been shown as dark and light coloured stars,

which correspond to the ratio of translational diffusion coefficients along the long axis and short axis of the nanohelix respectively. (d) Relative viscosity measured at a point close to the boundary as the mixing of two different fluids continued to occur. The ratio of the diffusion coefficients from passive measurements is also shown, similar to Fig. 2c. Note the spatial and temporal mapping experiments were done in two different microfluidic chambers at different times.

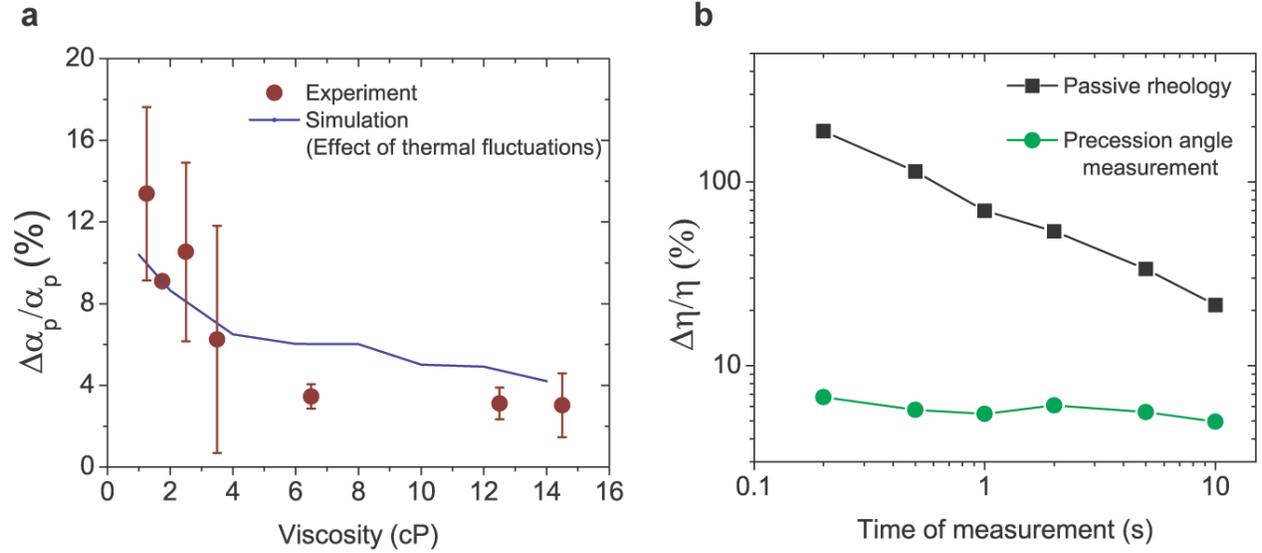

**Figure 3**: (a) Percentage error in measuring the precession angle as a function of the viscosity measured. The brown circles and error bars represent the values obtained from the experiment in which the nanohelices were moved in methyl cellulose of different viscosities, while the blue solid line is the simulated percentage error that considers the effect of thermal fluctuations only. (b) Experimentally obtained percentage errors in measuring the viscosity of methyl cellulose gel of viscosity 14 cP, for two different schemes of measurements. The black squares correspond to the error in estimating the viscosity using passive microrheology with a bead of diameter 1 µm. The green circles are the percentage errors in viscosity estimate in the same fluid using the methodology described here.

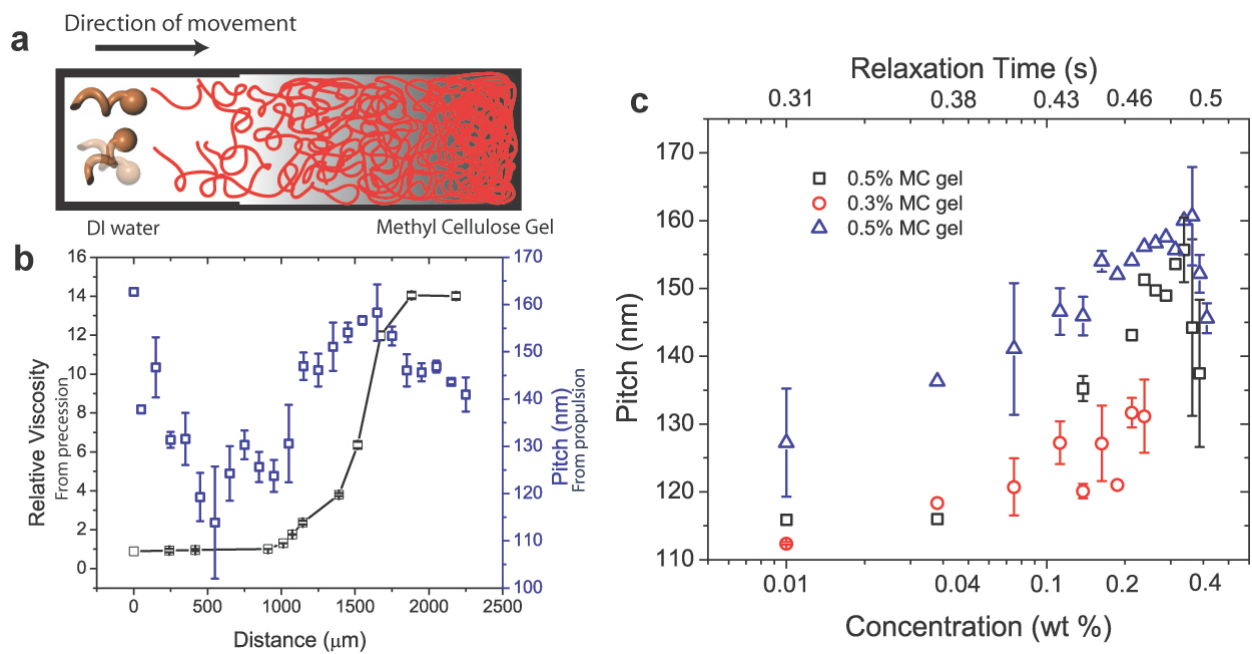

**Figure 4:** (a) Schematic of the measurement chamber containing DI water at one end and methyl cellulose gel (0.5 wt %) on the other side. The inset schematics show a helix driven from water towards the gel. The two different configurations of the same helix (propulsion (top) and precession (bottom)) were used to measure the pitch and viscosity respectively in a single experiment as a function of position. (b) Result of the simultaneous measurement of the local viscosity (black, left axis) and the local pitch (blue, right axis) as a function of position. The starting position (distance = 0 μm) corresponded to initial position in pure DI water, and subsequently the helices were moved towards the gel. The end position of the helix (distance = 2200 μm) corresponded to MC gel concentration of about 0.4%. (c) Results of multiple experiments on the dependence of the pitch of the helices on concentration of the MC gel. Different propellers and initial concentration of the MC gel were used in the different experiments. The variation of the pitch as a function of position was converted to the dependence of the pitch on the concentration of the gel (bottom axis) and the corresponding relaxation time (top axis) of the elastic medium. See text for detailed description of the experimental procedure.